\documentstyle[12pt]{article}

\def\L{{\cal L}}
\def\M{{\cal M}}
\def\Cinf{{C^{\infty}(\M)}}
\def\endproof{\hbox to \hsize{\hfil $\Box$}}
\def\endexample{\centerline{\vbox{\hrule width4in}}}
\begin{document}
\medskip
\centerline{{\Large\bf A Geometry for Multidimensional}}

\vspace{.2in}
\centerline{{\Large\bf Integrable Systems}}

\vspace{.2in}
\centerline{{\bf I.A.B.Strachan}}

\vspace{.1in}
\centerline{Dept. of Pure Mathematics and Statistics, University of Hull,}
\vspace{.1in}
\centerline{Hull, HU6 7RX, England
\footnote{e-mail: i.a.b.strachan@hull.ac.uk}.}

\vspace{.3in}
\centerline{{\bf Abstract}}

\vspace{.2in}
\small
\parbox{5.8in}{A deformed differential calculus is developed based on an 
associative $\star$-product. In two dimensions the Hamiltonian vector fields 
model the algebra of pseudo-differential operator, as used in the theory of 
integrable systems. Thus one obtains a geometric description of the operators.
A dual theory is also possible, based on a deformation of differential forms.
This calculus is applied to a number of multidimensional integrable systems,
such as the KP hierarchy, thus obtaining a geometrical description of these 
systems. The limit in which the deformation disappears corresponds to taking 
the dispersionless limit in these hierarchies.} 

\normalsize

\bigskip

\bigskip

\centerline{To appear, {\sl Journal of Geometry and Physics}.}

\bigskip

\section*{1. Introduction }

\bigskip

Is there a common structure behind all integrable systems? There are many 
different types of integrable systems; $(1+1)$ and $(2+1)$-dimensional 
evolution equations (such as the KdV and KP equations), chiral and harmonic map 
equations, integrable dynamical systems (such as the Halphen and Kovalevskaya 
top equations), integrable nonlinear ordinary differential equations (such as 
the Chazy and Painlev\'e equations), for example, and all these have various 
properties associated with their integrability (see, for example, [AC]). 
However, there is very little 
in the way of general theory, where the apparently disparate properties of 
various individual integrable systems could be understood in a consistent and 
coherent way. Indeed, there is not even a universal definition of what 
integrability actually is.

\bigskip

One idea, proposed by Ward [W], is that such system may all be obtained from the 
(anti)-self-dual Yang-Mills equations and their generalisation by a process of 
dimensional reduction. For example, the KdV, NLS, sine-Gordon and Liouville 
equations may all be obtains from the (anti)-self-dual Yang-Mills equations 
with $SL(2,{\bf C})$-gauge groups, the only difference being the choice of 
symmetry group and spacetime signature [W,MS]. The key idea is not so much the 
self-dual Yang-Mills equations themselves, but the existence of a Penrose 
transform for such fields. Under such a transform the fields \lq disappear\rq~ 
into the holomorphic geometry (certain holomorphic vector bundles over regions 
of an auxiliary complex manifold known as twistor space) [W77]. More prosaically, 
this provides a geometric framework for the Riemann-Hilbert problems used
in the construction of solutions to the duality equations.The existence of this 
transform has been conjectured to be behind the idea of integrability.

\bigskip

The paradigm (and the original example of such a transform) came from the 
(anti)-self-dual Vacuum equations [Pen]. The following formalism is due to 
Gindikin [G].
Consider the following system of first order equations depending on a parameter
$\tau=\{\tau_0,\tau_1\}\in{\bf C}^2\,:$

\begin{eqnarray*}
\omega^1(\tau) & = & \omega_0^1 \tau_0^k + \ldots + \omega_k^1 \tau_1^k \,, \\
\vdots &  & \vdots \\
\omega^{2p}(\tau) & = & \omega_0^{2p} \tau_0^k + \ldots + 
\omega_k^{2p} \tau_1^k \,
\end{eqnarray*}

\noindent where $\omega_j^i$ are 1-forms. Let ${\bf\Omega}^k(\tau)$ be the bundle 
of 2-forms

\[
{\bf\Omega}^k(\tau) = \omega^1(\tau)\wedge\omega^2(\tau) + \ldots +
\omega^{2p-1}(\tau) \wedge \omega^{2p}(\tau)
\]

\noindent satisfying the conditions:

\bigskip

$\bullet$ the $(p+1)^{\rm th}$ exterior power of ${\bf\Omega}^k$ vanishes;

\medskip

$\bullet$ the $p^{\rm th}$ exterior power of ${\bf\Omega}^k$ is non-degenerate;

\medskip

$\bullet$ $d {\bf\Omega}^k(\tau) = 0 \,.$

\bigskip

\noindent The bundle of forms then encodes the integrability of the original 
system. In the special case $k=1\,,p=1$ the metric defined by

\[
{\bf g} = \omega^1_0\, \omega^2_1 - \omega^1_1 \, \omega^2_0
\]

\noindent has vanishing Ricci tensor and (anti)-self-dual Weyl tensor. The form 
$\bf\Omega$ is related to various structures on the corresponding curved twistor space.

\bigskip

One major problem with this geometric approach to the understanding of 
integrable systems is to find how systems such as the KP equation

\[
(4 u_t - 12 u u_x - u_{xxx})_x = 3 u_{yy}
\]

\noindent fit into the scheme. There have been various attempts,
some erroneous, to do this, the problem stemming from how to give a geometrical 
description to the pseudo-differential operators used in the derivation of the 
KP equation and its hierarchy. However, though there are problems with the KP 
equation itself, these problem vanish in a particular limit (the dispersionless 
limit) of the KP equation and one obtains a geometric description of this 
limiting case. Explicitly, let

\begin{eqnarray*}
X & = & \epsilon x \,, \\
Y & = & \epsilon y \,, \\
T & = & \epsilon t \,, \\
U(X,Y,T) & = &u(x,y,t)\,,
\end{eqnarray*}

\noindent then the KP equation becomes, in the limit $\epsilon\rightarrow 0$ 
the dispersionless (or dKP equation):

\[
(4U_T - 12 U U_X)_X = 3 U_{YY}\,.
\]

\noindent This too is integrable, but the description does not use 
pseudo-differential operators but a more geometrical description in terms of 
vector fields and differential forms.

\bigskip

The central idea of this paper is the development of a deformed differential 
calculus based on an associative $\star$-product
and its application to the theory of integrable systems. One will 
obtains an elegant description of the KP hierarchy in terms of vector fields 
and differential forms rather than the more usual pseudo-differential operator 
formalism. The advantage of this approach will be two-fold: firstly, one 
retains a geometric description, so, for example, one can go to a dual description 
in terms of differential forms; secondly in the limit in which the deformation 
disappears one recovers the dKP equation directly without the need of the 
somewhat singular limit outlined above.

\bigskip

This deformed calculus will be derived in section 2 and used in section 3 where 
various examples of multidimensional integrable systems (an integrable 
deformation of the (anti)-self-dual vacuum equations, the KP hierarchy and the 
Toda hierarchy) will be studied. It will turn out the all these systems may be 
written in terms of a 2-form $\bf\Omega$ satisfying the equations

\begin{eqnarray*}
d{\bf\Omega} & = & 0 \,, \\
{\bf\Omega}\wedge{\bf\Omega} & = & 0 
\end{eqnarray*}

\noindent in analogy to Gindikin's bundle of forms. In section 4 the geometry 
of the KP hierarchy will be examined in more detail. This work raises a number 
of further questions, some of which are outlined in section 5.

\bigskip

\section*{2. Deformed Differential Geometry}

\bigskip

A Poisson manifold $\M$ is endowed with a bilinear skew-symmetric Poisson 
bracket defined, for $u,v\in\Cinf\,,$ by

\begin{equation}
\{u,v\}_{PB} = \omega^{ij} {\partial u\over\partial x^i}{\partial v\over
\partial x^j}\,,
\label{eq:PB}
\end{equation}

\noindent and with the additional property that

\[
\{\{f,g\}_{PB},h\}_{PB} + \{\{g,h\}_{PB},f\}_{PB} + 
\{\{h,f\}_{PB},g\}_{PB} = 0 \,,
\]

\noindent this being known as the Jacobi identity. Further, it will be assumed 
that $\M$ is a symplectic manifold, that is a Poisson 
manifold for which the matrix $\omega^{ij}$ is of maximal rank. It follows that 
the dimension of $\M$ must be even, so

\[
{\rm dim~}(\M) = 2N
\]

\noindent for some integer $N\,.$ It will be assumed that $\omega^{ij}$ is 
constant and that a basis has been chosen in which

\[
\omega^{ij}=\left( \begin{array}{cc} 0 & I_N \\ -I_N & 0 \end{array} \right) 
\,.
\]

\bigskip

This structure may be used to define a deformation of the above Poisson 
bracket. For $u,v\in\Cinf$ one defines a new product

\[
u \star v = \exp\Big({\kappa\over 2} \, \omega^{ij} 
{\partial\phantom{x^i}\over\partial x^i}{\partial\phantom{x^j}\over\partial
{\tilde x}^j}\Big) u({\bf x}) v({\bf{\tilde x}})
\Big\vert_{{\bf x}={\bf{\tilde x}}}
\]

\noindent or, on expanding the exponential

\begin{equation}
u \star v = \sum_{s=0}^{\infty} {\kappa^s \over 2^s s!}
\omega^{i_1j_1} \ldots \omega^{i_sj_s}
{\partial^s u \over\partial x^{i_1} \ldots \partial x^{i_s}}
{\partial^s v \over\partial x^{j_1} \ldots \partial x^{j_s}}\,.
\label{eq:star}
\end{equation}

\noindent With this the deformed, or Moyal bracket, is defined as [Mo]

\begin{equation}
\{u,v\} = {u \star v - v \star u \over \kappa}\,.
\label{eq:MB}
\end{equation}

\bigskip

\noindent {\bf Lemma 1}

\medskip

For constants $c$ and functions $u,v\in \Cinf\,,$

\medskip

\indent (a) $lim_{\kappa\rightarrow 0} \, u \star v = uv\,,$

\medskip

\indent (b) $c \star u = cu\,,$

\medskip

\indent (c) $\star$ is associative\,,

\medskip

\indent (d) $lim_{\kappa\rightarrow 0} \, \{u,v\} = \{u,v\}_{PB}\,,$

\medskip

\indent (e) $\{u,v\}$ is bilinear, skew-symmetric and satisfies the Jacobi 
identity.

\bigskip

\noindent {\bf Proof}

\bigskip

Straightforward from definitions (\ref{eq:PB})~and (\ref{eq:star}). 
Note that the Jacobi identity 
follows from the associativity of the $\star$-product.

\bigskip

\endproof

\bigskip

The original motivation for the introduction of such a bracket came from a 
description of quantum mechanics using phase space variables [Mo]. Here
$\kappa$ is replaced by 
$-i\hbar\,,$ and $\hbar\rightarrow 0$ corresponds to taking the classical 
limit. 

\bigskip

It is will necessary to introduce another product. For $u,v \in\Cinf$ define

\begin{equation}
u \circ v = \sum_{s=0}^{\infty} {\kappa^{2s} \over 2^{2s}(2s+1)!}
\omega^{i_1j_1} \ldots \omega^{i_{2s}j_{2s}}
{\partial^{2s} u \over\partial x^{i_1} \ldots \partial x^{i_{2s}}}
{\partial^{2s} v \over\partial x^{j_1} \ldots \partial x^{j_{2s}}}\,.
\label{eq:circ}
\end{equation}

\bigskip

\noindent {\bf Lemma 2}

\bigskip

For constants $c$ and function $u,v\in\Cinf\,,$

\medskip

\indent (a) $u \circ v = v \circ u \,,$

\medskip

\indent (b) $c \circ u = cu\,$

\medskip

\indent (c) $lim_{\kappa\rightarrow 0} \, u \circ v = uv\,,$

\medskip

\indent (d) $\circ$ is {\sl not} associative,

\medskip

\indent (e) $2{d (\kappa \, u\circ v) / d\kappa} = u\star v + v\star u\,.$

\bigskip

\noindent {\bf Proof}

\bigskip

Again, these results follow from definitions (\ref{eq:star})~and 
(\ref{eq:circ}).
                                                      
\bigskip

\endproof

\bigskip

With this $\circ$-product a deformed, or quantum, differential calculus will be 
constructed. A similar calculus has recently been constructed by Fedosov [Fe], this 
using the $\star$-product rather than the $\circ$-product. The reason for the 
introduction of the new product will become apparent later (the motivation 
coming from the application of this calculus to multidimensional integrable 
systems) and rests on the following result:

\bigskip

\noindent {\bf Proposition 3}

\bigskip

For $u,v \in \Cinf\,,$

\[
\omega^{rs} {\partial u\over \partial x^r}\circ{\partial v\over \partial x^s}
= \{u,v\}\,.
\]

\bigskip

\noindent {\bf Proof}

\bigskip

This follows from the definitions ({\ref{eq:MB}})~and ({\ref{eq:circ}}):

\begin{eqnarray*}
\omega^{rs} {\partial u\over \partial x^r}\circ{\partial v\over \partial x^s}
& = &                                                                         
\phantom{{2\over\kappa}}
\sum_{s=0}^{\infty} {\kappa^{2s}\over 2^{2s}(2s+1)!}
\omega^{i_1j_1} \ldots \omega^{i_{2s}j_{2s}} \omega^{rs}
{\partial^{2s+1} u \over\partial x^{i_1} 
\ldots \partial x^{i_{2s}}\partial x^r}
{\partial^{2s+1} v \over\partial x^{j_1} 
\ldots \partial x^{j_{2s}}\partial x^s}      \,.  \\
& = &
{2\over\kappa}\sum_{s=0}^{\infty} {\kappa^{2s+1} \over 2^{2s+1}(2s+1)!}
\omega^{i_1j_1} \ldots \omega^{i_{2s+1}j_{2s+1}}
{\partial^{2s+1} u \over\partial x^{i_1} \ldots \partial x^{i_{2s+1}}}
{\partial^{2s+1} v \over\partial x^{j_1} \ldots \partial x^{j_{2s+1}}} \,, \\ 
& = &
\phantom{{2\over\kappa}} \Big( {u \star v-v \star u \over \kappa} \Big) \,, \\
& = &
\phantom{{2\over\kappa}}\{u,v\}\,.
\end{eqnarray*}

\noindent In the limit $\kappa\rightarrow 0$ this reduces to the definition 
({\ref{eq:PB}}).

\bigskip

\endproof

\bigskip

In the simplest case ($N=1$) these formulae may be easily rewritten using the 
explicit form of $\omega^{ij}\,,$ coordinates $x^1=x\,,x^2=y\,:$

\begin{eqnarray}
u \star v & = & \sum_{s=0}^\infty {\kappa^s\over 2^s s!} \sum_{j=0}^s (-1)^j
\left( \begin{array}{c} s \\ j \end{array} \right)
{\partial^{s-j}_x}
{\partial^{j}_y} u \,
{\partial^{j}_x}
{\partial^{s-j}_y} v \,, 
\label{eq:2dcasea} \\
u \circ v & = & \sum_{s=0}^\infty {\kappa^{2s}\over 2^{2s+1}(2s+1)!} 
\sum_{j=0}^{2s} (-1)^j
\left( \begin{array}{c} 2s \\ j \end{array} \right)
{\partial^{2s-j}_x}
{\partial^{j}_y} u \,
{\partial^{j}_x}
{\partial^{2s-j}_y} v \,, 
\label{eq:2dcaseb} \\
\{u,v\} & = & \sum_{s=0}^\infty {\kappa^{2s+1}\over 2^{2s+1}(2s+1)!} 
\sum_{j=0}^{2s+1} (-1)^j
\left( \begin{array}{c} 2s+1 \\ j \end{array} \right)
{\partial^{2s+1-j}_x}
{\partial^{j}_y} u \,
{\partial^{j}_x}
{\partial^{2s+1-j}_y} v 
\label{eq:2dcasec}
\end{eqnarray}

\noindent for functions $u(x,y)\,,v(x,y)\in\Cinf\,.$

\bigskip

\noindent {\bf Example 1}

\bigskip

Let $\M=T^2\,,$ the 2-torus. Functions on $T^2$ may be expanded in terms of 
basis functions

\[
e_{\bf m} = \exp i(m_1 x + m_2 y)\,.
\]

\noindent With these one obtains from (\ref{eq:2dcasea})-(\ref{eq:2dcasec}):

\begin{eqnarray*}
e_{\bf m} \star e_{\bf n} & = & \phantom{2}\exp({\kappa\over 2} {\bf m}\times{\bf n})\, 
e_{{\bf m}+{\bf n}} \,, \\
\{e_{\bf m},e_{\bf n}\} & = & \phantom{2}\sinh{\kappa\over 2}
({\bf m}\times{\bf n}) \, e_{{\bf m}+{\bf n}} \,, \\
e_{\bf m} \circ e_{\bf n} & = & {2}{\sinh{\kappa\over 2}({\bf m}\times{\bf n})\over
\kappa ({\bf m} \times {\bf n}) } \, e_{{\bf m}+{\bf n}} \,, 
\end{eqnarray*}

\noindent where ${\bf m}\times{\bf n} = m_2 n_1 - m_1 n_2\,.$

\bigskip

\endexample

\bigskip

Given such a symplectic manifold and product one may define tangent and 
cotangent bundles $T\M$ and $T^*\M\,,$ the inner product between basis
elements $\partial\phantom{x^i}\over\partial x^i$ and $dx^j$ being given
by

\[
\langle {\partial\phantom{x^i}\over\partial x^i},dx^k\rangle = \delta^j_i\,.
\]

\noindent The first difference is the formula for the inner product between 
general elements $X\in T_p\M$ and $\omega\in T^*_p\M\,,$

\begin{eqnarray*}
\langle X,\omega \rangle & = & \langle X^i 
{\partial\phantom{x^i}\over\partial x^i},\omega_j dx^j \rangle \,, \\
& = & X^i \circ \omega_j
\langle {\partial\phantom{x^i}\over\partial x^i},dx^k\rangle \,, \\
& = & X^i \circ \omega_i\,,
\end{eqnarray*}

\noindent i.e. the multiplication being done with the $\circ$-product. 
Similarly, given a vector field $X$ and function $f$ one defines

\begin{equation}
Xf = X^i \circ {\partial f\over \partial x^i}\,,
\label{eq:vector}
\end{equation}

\noindent again using the $\circ$-product.
The general procedure should already be apparent: the only change to the 
standard, or undeformed, theory is when objects are combined, this being done 
with the $\circ$-product. Thus in the $\kappa\rightarrow 0$ limit the standard 
theory is recovered. One may extend this new calculus to general tensor fields.
However of more interest is the extension to an exterior differential calculus.

\bigskip

The $d$-operator, which maps $r$-forms to $(r+1)$-forms is defined as normal.
For example, given a $0$-form (i.e. a function) the $1$-form $df$ is defined by
the relation

\[
\langle X, df \rangle = Xf
\]

\noindent for all vector fields $X\,.$ From this follows the formula

\[
df={\partial f\over\partial x^i} dx^i\,.
\]

\noindent The wedge product combines forms and so this will be done using the
$\circ$-product. Explicitly, if

\begin{eqnarray*}
{\bf A} & = & A_{i_1\ldots i_p} dx^{i_1}\wedge\ldots\wedge dx^{i_p}\,,\\
{\bf B} & = & B_{j_1\ldots j_q} dx^{j_1}\wedge\ldots\wedge dx^{j_q}   
\end{eqnarray*}

\noindent then   

\[
{\bf A}\wedge{\bf B} = A_{[i_1\ldots i_p}\circ B_{j_1\ldots j_q]}
dx^{i_1}\wedge\ldots\wedge dx^{i_p} \wedge dx^{j_1}\wedge\ldots\wedge 
dx^{j_q}\,.
\]

\bigskip

\noindent {\bf Example 2}

\bigskip
   
Suppose ${\rm dim~}\M=2$ (i.e. $N=1$). Then for functions $f(x,y)\,,g(x,y)\in\Cinf\,,$

\begin{eqnarray*}
df & = & f_x dx + f_y dy \,, \\
dg & = & g_x dx + g_y dy \,,
\end{eqnarray*}

\noindent and hence

\begin{eqnarray*}
df\wedge dg & = & (f_x \circ g_y - f_y \circ g_x ) \, dx \wedge dy \,, \\
            & = & \{f,g\} \, dx \wedge dy\,.
\end{eqnarray*}

\noindent Note that this uses the symmetry property of the $\circ$-product.
Care must be taken in higher dimensions since, as the $\circ$-product is not 
associative, ${\bf A}\wedge ({\bf B}\wedge{\bf C}) \neq ({\bf A}\wedge{\bf B})
\wedge{\bf C}$ for arbitrary forms ${\bf A}\,,{\bf B}$ and ${\bf C}\,.$

\bigskip

\endexample

\bigskip

Having defined an exterior differential calculus one may define another 
intrinsic differential object, namely a Lie derivative $\L_X$ corresponding 
to some vector field $X\in T\M\,.$ On functions 

\[
\L_X f = Xf
\]

\noindent and on vector fields

\[
(\L_X Y)^i = X^j \circ {\partial Y^i\over \partial x^j} - Y^j \circ
{\partial X^i \over \partial x^j}\,.
\]

\noindent This will also be written as $\L_X Y = [X,Y]\,,$ and called the 
commutator of two vector fields. Using the symmetry of the $\circ$-product it 
follows 
that the commutator is antisymmetric. One may extend the definition to more 
general objects such as tensor fields in such a way that the theory is 
consistent. For example, for any $p$-form $\omega$ and vector field $X\,,$

\[
d(\L_X \omega) = \L_X (d\omega)\,.
\]

Normally one has the relations

\begin{eqnarray*}
{[X,Y]}f - X(Yf) + Y(Xf) & = & 0 \,,  \\
{[{[X,Y]},Z]}+{[{[Y,Z]},X]}+{[{[Z,Y]},X]} & = & 0 \,.
\end{eqnarray*}

\noindent The proof of these results uses the 
associativity of normal (i.e. undeformed) multiplication, and so do not 
hold for the deformed definitions based on the non-associative $\circ$-product.
However for an important class of vector fields these results do hold. 
Given a function $f\in\Cinf$ (the Hamiltonian) the corresponding 
Hamiltonian vector field $X_f$ is defined by

\begin{equation}
X_f=\omega^{ij} {\partial f\over\partial x^i} {\partial\phantom{x^j}\over 
\partial x^j}\,.
\label{eq:hamvector}
\end{equation}

\noindent Strictly speaking these are local Hamiltonian vector fields, 
Hamiltonian vector fields having to be defined globally on $\M\,.$

\bigskip

\noindent {\bf Lemma 4}

\bigskip

For functions $f,g,h\in\Cinf$ and corresponding Hamiltonian fields $X_f,X_g$ 
and $X_h\,,$

\bigskip

\indent (a) $X_f h = \{f,h\}\,,$

\medskip

\indent (b) $[X_f,X_g]=X_{\{f,g\}}\,,$

\medskip

\indent (c) $[X_f,X_g] h = X_f(X_g h) - X_g (X_f h) \,,$

\medskip

\indent (d) $[[X_f,X_g],X_h]+[[X_g,X_h],X_f]+[[X_h,X_f],X_g] = 0 \,.$

\bigskip

\noindent {\bf Proof}

\bigskip

(a) This follows from definitions (\ref{eq:vector}), (\ref{eq:hamvector}) and
Proposition 3.

\medskip

(b)

\begin{eqnarray*}
[X_f,X_g]^i & = & X_f^j \circ {\partial X_g^i\over \partial x^j} -
                  X_g^j \circ {\partial X_f^i\over \partial x^j}\,, \\
& = & \omega^{kj} {\partial f\over \partial x^k} \circ 
{\partial\phantom{x^j}\over \partial x^j} \Big( \omega^{ri}
{\partial g\over \partial x^r}\Big) -
\omega^{kj} {\partial g\over \partial x^k} \circ 
{\partial\phantom{x^j}\over \partial x^j} \Big( \omega^{ri}
{\partial f\over \partial x^r}\Big) \,, \\
& = & \omega^{ri} {\partial\phantom{x^r}\over \partial x^r}
\Big( \omega^{kj} {\partial f\over \partial x^k}\circ{\partial g\over \partial 
x^j} \Big)\,, \\
& = & X_{\{f,g\}}^i\,.
\end{eqnarray*}

\noindent This uses the antisymmetry of $\omega^{ij}$ and the symmetry of the 
$\circ$-product.

\medskip

(c)

\begin{eqnarray*}
[X_f,X_g] h & = & X_{\{f,g\}} h \,, \\
& = & \{\{f,g\}h\} \,,  \\
& = & \{f,\{g,h\}\} - \{g,\{f,h\}\} \,, \\
& = & X_f(X_g h) - X_g(X_f h) \,.
\end{eqnarray*}

\noindent This uses result (b), the Jacobi identity for the Moyal bracket 
and the antisymmetry of the Moyal bracket (Lemma 1).

\medskip

(d) This follows from the Jacobi identity for the Moyal bracket (Lemma 1).

\bigskip

\endproof

\bigskip

\noindent {\bf Example 3}

\bigskip

For $2$-dimensional manifolds $\M^2$ these Hamiltonian vector fields generate 
the Lie algebra of area preserving diffeomorphisms of the manifold where 
the area 
element is $dx \wedge dy$ and the composition of two Hamiltonian vector fields 
is defined to be the Lie bracket of these fields. Explicitly, the field
$X_f$ generates the infinitesimal transformation $x\rightarrow x-\epsilon 
f_y\,,$ $y\rightarrow y+\epsilon f_x\,.$ This Lie algebra will be denoted by
$sdiff_\kappa(\M^2)\,.$

\bigskip

\endexample

\bigskip

The differential objects constructed have been intrinsic to the manifold. One 
should be able to introduce a connection on $\M$ and define a covariant 
differentiation and hence curvature. A similar programme has been carried out 
using the $\star$-product by Vasiliev [V]. However for the application of this calculus 
to the theory of integrable systems such a structure will not be required.

\bigskip

This $\star$-product is essentially unique. For a product

\[
f\star g = f g + \sum_{r=1}^\infty \kappa^r Q_r(f,g)
\]

\noindent (where the $Q_r$ are bilinear differential operators) to be 
associative places strong restrictions on the type of higher-order terms that 
may be added. Further, the requirement that the bracket defined by
$\{f,g\}^{\prime} = (f\star g - g\star f)/\kappa$ should reduce to the Poisson bracket 
in the $\kappa \rightarrow 0 $ limit introduces further restrictions and from 
these considerations follow various results on the uniqueness of the Moyal 
bracket [A,BFFLS,Fl].
However these uniqueness results  
only state that any such deformations are equivalent to the Moyal bracket; there 
are apparently different structures which, after various changes of variable,
become the Moyal bracket. For example one may define the following associative 
$\star$-product (in the $N=1$ case)

\begin{equation}
f \star g = \sum_{s=0}^\infty {\kappa^s\over s!}
{\partial^s f \over \partial x^s}{\partial^s g \over \partial y^s}
\label{eq:mstar}
\end{equation}

\noindent This too defines a bracket

\begin{equation}
\{f,g\}^\prime = {f\star g - g \star f\over \kappa}
\label{eq:KMB}
\end{equation}

\noindent which reduces to the standard Poisson bracket in the 
$\kappa\rightarrow 0 $ limit. This new bracket will to called the 
Kupershmidt-Manin bracket [K,Ma]. As above, one may define a corresponding 
$\circ$-product

\begin{equation}
f\circ g = \sum_{s=0}^\infty {\kappa^s\over (s+1)!} \sum_{m=0}^s
\partial_x^{s-m}\partial_y^m f \, \partial_y^{s-m}\partial_x^m g\,,
\label{eq:mcirc}
\end{equation}

\noindent so that

\[
\omega^{rs} 
{\partial u \over \partial x^r} \circ
{\partial v \over \partial x^s}  = \{f,g\}^\prime
\]

\noindent and hence an equivalent deformed differential geometry based on these 
new structures. The form of the $\star$-product is somewhat simplier then that 
given by (\ref{eq:2dcasea}), though the dependence on the sympletic structure of $\M$ is less 
transparent. The importance of these new products comes from their relationship 
to the algebra of pseudo-differential operators.

\bigskip

A pseudo-differential operator $P$ is an operator of the form

\[
P=\sum_{j=-\infty}^{\rm finite} a_j(x) \partial^j
\]

\noindent where the multiplication of two such operators uses the generalised 
Leibnitz rule

\[
\partial^m a = a \partial^m + \sum_{k=0}^\infty
{m(m-1) \ldots (m-k-1) \over k!} \partial^k a \,\, \partial^{m-k}\,.
\]

\noindent The set of such operators will be denoted by ${\cal P}\,.$
The symbol of a pseudo-differential operator is a function of two variables 
defined by

\[
sym( \sum_{j=-\infty}^{\rm finite} a_j(x) \partial^j) =
     \sum_{j=-\infty}^{\rm finite} a_j(x) y^j\,.
\]

\noindent It has the important property that for all $P\,,Q \in {\cal P}$

\[
sym(PQ) = sym(P) \star_{\kappa=1} sym(Q)
\]

\noindent where $\star_{\kappa=1}$ denotes the $\star$-product (\ref{eq:mstar}) 
evaluated at $\kappa=1\,.$ It follows from this that

\[
sym([P,Q]) = \{sym(P),sym(Q) \}^{\prime}_{\kappa=1}
\]

\noindent where $[P,Q]=PQ - QP\,.$ Thus one may replace pseudo-differential 
operators and its corresponding algebra by functions of two-variables
where the composition of two functions is done with the Kupershmidt-Manin 
bracket evaluated at $\kappa=1\,.$ 
More details of these algebraic properties may be found in [F-FMR].
Using the ideas developed above one may give 
these pseudo-differential operators a geometrical interpretation.

\bigskip

\noindent {\bf Theorem}

\bigskip

Let ${\cal H}$ be the space of Hamiltonian vector fields (where $N=1$) whose
Hamiltonians have Laurent expansions in the variable $y\,,$

\[
{\cal H} = \{ X_f \, : f=\sum_{j=-\infty}^{\rm finite} a_j(x) y^j \, \} \,.
\]

\noindent Then the map

\[
\iota \, : {\cal P}/{\{\rm constants\}} \rightarrow {\cal H}
\]

\noindent given by

\[
\iota(P) = X_{sym(P)}
\]

\noindent is an isomorphism. Moreover

\begin{eqnarray*}
\iota([P,Q]) & = & [ X_{sym(P)}, X_{sym(Q)} ]   \\
             & = & X_{ \{sym(P),sym(Q)\}^{\prime}_{\kappa=1} } 
\end{eqnarray*}

\noindent where the Lie bracket of vector fields is evaluated using the product
$\circ_{\kappa=1}$ given by (\ref{eq:mcirc})~evaluated at $\kappa=1\,.$

\bigskip

\noindent {\bf Proof}

\bigskip

Straightforward. Given a Hamiltonian vector field one can construct the 
corresponding Hamiltonian (up to a constant) and hence a pseudo-differential 
operator whose symbol is the Hamiltonian. Conversely, $\iota(P+c)=\iota(P)$ 
for all $P\in{\cal P}\,.$ The last part of the theorem follows 
from the properties of the symbol map.

\bigskip

\endproof

\bigskip

The set of Hamiltonian vector fields clearly forms a Lie algebra under the 
composition defined by the Lie bracket. One may define the adjoint 
representation as follows. For functions $f\,,g\,,F\in\Cinf$ define

\begin{eqnarray*}
ad(f) \, g & = & \{ f , g \} \,, \\
Ad(F) \, g & = & F \star g \star F^{-1} \,,
\end{eqnarray*}

\noindent the connection between the two coming from the deformed exponential

\[
\exp_\kappa f = 1 + \sum_{n=0}^\infty {1\over n!} \,
{\underbrace{f \star \ldots \star f}_{\rm n-terms}}\,.
\]

\noindent So, if $F=\exp_\kappa f\,,$

\[ 
Ad(F) \, g = \exp_\kappa(ad(f)) \, g
\]

\noindent (this uses the Baker-Campbell-Hausdorff formula). On vector fields,

\begin{eqnarray*}
ad(X_f) \, X_g & = & X_{\{f,g\}} \,, \\
Ad(F) \, X_g & = & X_{F\star g \star F^{-1}} \,.
\end{eqnarray*}

\noindent These will be used in section 4 to describe the dressing properties 
of the KP hierarchy.

\bigskip

The residue of a pseudo-differential operator $P=\sum a_n\partial^n$ is defined 
by

\[
res(P) = a_{-1}\,.
\]

\noindent It follows that

\begin{eqnarray*}
res(P) & = & res(sym(P)) \,, \\
       & = & {1\over 2\pi i} \oint sym(P) d y
\end{eqnarray*}

\noindent where the residue of the function $sym(P)$ is the normal residue, 
regarding $sym(P)$ as a complex function of $y\,.$ The residue has many uses, 
in particular in the study of the Hamiltonian properties of integrable systems.

\bigskip

\section*{3. Applications to Integrable Systems}

\bigskip

In this section a number of multidimensional integrable systems will be studied 
using the geometric structures developed in the last section. It will be shown 
that these systems may all be written in terms of a 2-form $\bf\Omega$ which 
satisfies the equations

\begin{eqnarray*}
d {\bf\Omega} & = & 0 \,, \\
{\bf\Omega}\wedge{\bf\Omega} & = & 0 \,.
\label{eq:universal}
\end{eqnarray*}

\noindent These equations encode the integrability conditions for these systems 
in an elegant geometric manner.

\bigskip

Let $\M$ be a sympletic manifold with some associated $\star$-product. In 
applications one will require an extended manifold

\[
{\widetilde\M} = \M \oplus {\cal T}
\]

\noindent where $\cal T$ consists of an extra set of coordinates (for example 
the \lq times\rq~in a hierarchy of evolution equations). The manifold $\M$ may 
be thought of as a phase space and in the applications considered here this 
will be two-dimensional. The $\star$-product extends to a product on 
$\widetilde\M$ by

\[
u({\bf x},{\bf t}) \star v({\bf x},{\bf t}) = \exp\Big(\kappa \, \omega^{ij} 
{\partial\phantom{x^i}\over\partial x^i}{\partial\phantom{x^j}\over\partial
{\tilde x}^j}\Big) u({\bf x},{\bf t}) v({\bf{\tilde x}},{\bf t})
\Big\vert_{{\bf x}={\bf{\tilde x}}}
\]

\noindent (where ${\bf x}=\{x^i\}$ are coordinates on $\M$ and $\bf t$ are 
coordinates on $\cal T$), that is, the dependence on the coordinates on         
$\cal T$ is ignored. The differential calculus outlinded in section 2 
similiarly extends to the manifold $\widetilde\M\,.$ One difference is that 
$X_f$ will refer to a Hamiltonian vector field on $\M$ whose Hamiltonian 
function depends on the coordinates on $\widetilde\M\,,$

\[
X_f = \omega^{ij} {\partial f({\bf x},{\bf t})\over\partial x^i}
{\partial\phantom{x^j}\over\partial x^j}\,,
\]

\noindent i.e. a time dependent Hamiltonian vector field on $\M\,,$ where the 
\lq times\rq~ are the coordinates on ${\cal T}\,.$

\bigskip

\subsection*{3.1 The Anti-Self-Dual Vacuum Equations}

\bigskip

The anti-self-dual vacuum equations govern the behaviour of complex 4-metrics 
of signature $(+,+,+,+)$ whose Ricci tensor is zero and whose Weyl tensor is 
anti-self-dual. Since these curvature conditions are invariant under changes of 
coordinates there are many ways to write these equations. On particular form of 
the equations uses the fact that such metrics are automatically K\"ahler and so 
may be written in terms of a single scalar function $\Omega\,,$ the K\"ahler 
potential. The curvature conditions then give the equation governing the 
potential (known as Plebanski's $1^{\rm st}$ Heavenly equation [Pl]):

\begin{equation}
{\partial^2\Omega\over\partial x \partial {\tilde x}}
{\partial^2\Omega\over\partial y \partial {\tilde y}} -
{\partial^2\Omega\over\partial x \partial {\tilde y}}
{\partial^2\Omega\over\partial y \partial {\tilde x}} = 1 \,.
\label{eq:Plebanski}
\end{equation}

\noindent The corresponding anti-self-dual Ricci-flat metric is

\begin{equation}
{\bf g}(\Omega) = 
{\partial^2\Omega\over\partial x^i \partial {\tilde x^j}} dx^i d{\tilde x}^j\,,
\quad\quad{\tilde x}^i = {\tilde x}\,,{\tilde y}\,,\quad x^j = x\,,y\,.
\label{eq:metric}
\end{equation}

\noindent This equation can, in principle, be solved using a Penrose transform 
- the original non-linear graviton construction [Pen]. Although not realised at the 
time, the existence of such a transform makes (\ref{eq:Plebanski}) into a 
completely integrable system, an important, and rare, example of a 
multidimensional integrable system. As such it has all the properties one would 
expect of such a system, an infinite number of conservation laws [S93] and an 
associated hierarchy [S95b], for example. A Lax pair for this equation was derived by
[NPT] and later interpreted by Park [Pa] as that for a 2-dimensional topological 
chiral model with gauge potentials in the infinite dimensional Lie algebra
$sdiff(\M^2)$ for some two dimensional manifold $\M\,.$

\bigskip

The equation may be written as

\[
\{ \Omega_x , \Omega_y \}_{\rm PB} = 1  \,,
\]

\noindent where the Poisson bracket is defined by

\[
\{ f , g \}_{\rm PB} = f_{\tilde x} \, g_{\tilde y} - f_{\tilde y}\,  g_{\tilde x}\,.
\]

\noindent The equation that will be studied in this section is an integrable 
deformation of this equation, where the Poisson bracket has been replaced by 
the Moyal bracket (\ref{eq:MB})

\begin{equation}
\{ \Omega_x , \Omega_y \} = 1 \,.
\label{eq:mPlebanski}
\end{equation}

\noindent The space $\M$ will have coordinates $\{ {\tilde x},{\tilde y} \}$ 
(with composition using the products (\ref{eq:star}) and (\ref{eq:circ})) and 
the space $\cal T$ will be taken to be ${\bf R}^2$ (or possibly ${\bf C}^2$) 
with coordinates $\{x,y\}\,.$ This deformed system (\ref{eq:mPlebanski}) will 
be studied using the deformed calculus developed in section 2. It will turn out 
the it shares many of the features and properties of the undeformed system 
(\ref{eq:Plebanski}). 

\bigskip

Let ${\cal U}$ and ${\cal V}$ be the following vector fields on $T{\widetilde\M}$

\begin{eqnarray*}
{\cal U} & = & \lambda {\partial\phantom{x}\over\partial x} + X_f \,, \\
{\cal V} & = & \lambda {\partial\phantom{y}\over\partial y} + X_g \,,
\end{eqnarray*}

\noindent where $\lambda\in{\bf CP}^1$ is a constant known as the spectral 
parameter. The system of equations for the function 
$\psi\in C^{\infty}(\widetilde\M)$

\begin{eqnarray*}
{\cal U} (\psi) & = & 0 \,, \\
{\cal V} (\psi) & = & 0 \,
\end{eqnarray*}

\noindent (or, equivalently,

\begin{eqnarray*}
\lambda \psi_x + \{ f , \psi \} & = & 0 \,, \\
\lambda \psi_y + \{ g , \psi \} & = & 0 \,)\,, 
\end{eqnarray*}

\noindent is overdetermined unless the integrability condition

\[
[ {\cal U} , {\cal V} ] = 0
\]

\noindent holds. Here $[ \, , \, ]$ is the Lie bracket of vector fields. 
If these equations are satisfied, then one has two independent solutions 
$L$ and $M$ for $\psi$ which satisfy the equation $\{L,M\}=1\,.$ Note that the 
above equations may be written in the following ways:

\begin{eqnarray*}
{\cal U} (L) = {\cal U} (M) & = & 0 \,, \\
{\cal V} (L) = {\cal V} (M) & = & 0 \,,
\end{eqnarray*}

\noindent or

\begin{eqnarray*}
\langle {\cal U}, dL \rangle = \langle {\cal U}, dM \rangle & = & 0 \,, \\
\langle {\cal V}, dL \rangle = \langle {\cal V}, dM \rangle & = & 0 \,.
\end{eqnarray*}

\noindent The above integrability conditions are satisfied if the 
functions $f$ and $g$ satisfy the equations

\begin{eqnarray*}
f_y - g_x & = & 0 \,, \\
\{ f , g \} & = & 1 \,.
\end{eqnarray*}

\noindent The first equation implies the existence of a scalar function 
$\Omega$ such that $f=\Omega_x\,,g=\Omega_y$ and with these the second equation 
becomes the deformed Plebanski equation (\ref{eq:mPlebanski}). This shows that 
this may be interpreted as a 2-dimensional chiral model with gauge potentials 
in the Lie algebra $sdiff_{\kappa}(\M^2)\,.$

\bigskip

In [S92] the vector fields ${\cal U}$ and ${\cal V}$ were interpretated 
as operators:

\begin{eqnarray*}
{\cal U}& = & \lambda \partial_x +
\sum_{s=0}^\infty {\kappa^{2s+1}\over 2^{2s+1}(2s+1)!} 
\sum_{j=0}^{2s+1} (-1)^j
\left( \begin{array}{c} 2s+1 \\ j \end{array} \right)
{\partial^{2s+1-j}_{\tilde x}}
{\partial^{j}_{\tilde y}} f  \,
{\partial^{j}_{\tilde x}}
{\partial^{2s+1-j}_{\tilde y}} \,, \\ 
{\cal V}& = & \lambda \partial_y +
\sum_{s=0}^\infty {\kappa^{2s+1}\over 2^{2s+1}(2s+1)!} 
\sum_{j=0}^{2s+1} (-1)^j
\left( \begin{array}{c} 2s+1 \\ j \end{array} \right)
{\partial^{2s+1-j}_{\tilde x}}
{\partial^{j}_{\tilde y}} g \,
{\partial^{j}_{\tilde x}}
{\partial^{2s+1-j}_{\tilde y}} \,. 
\end{eqnarray*}

\noindent The geometrical approach used here is much simplier, and the 
manipulations using the $\circ$-product which lead to equation 
(\ref{eq:mPlebanski})~are almost transparent, deviating very little from the 
undeformed calculation which leads to equation (\ref{eq:Plebanski}) (to achieve 
such a result was one of the original motivations in the development of the 
deformed calculus). Another advantage of the geometrical over the operator 
based approach is that one can go over to a dual description in terms of 
differential forms on $T^{*}(\widetilde\M)\,.$

\bigskip

Let ${\bf\Omega}$ be the 2-form

\[
{\bf\Omega} = dx \wedge dy + \lambda 
(\Omega_{x{\tilde{x}}} dx \wedge d{\tilde{x}} + 
\Omega_{x{\tilde{y}}} dx \wedge d{\tilde{y}} + 
\Omega_{y{\tilde{x}}} dy \wedge d{\tilde{x}} + 
\Omega_{y{\tilde{y}}} dy \wedge d{\tilde{y}})+ 
\lambda^2 d{\tilde{x}}\wedge d{\tilde{y}} \,.
\]

\noindent This clearly satisfies the condition $d{\bf\Omega}=0\,,$ and in 
addition

\begin{eqnarray*}
{\bf\Omega}\wedge{\bf\Omega} & = & \lambda^2
(\Omega_{x{\tilde x}} \circ \Omega_{y{\tilde y}} - \Omega_{x{\tilde y}} \circ
\Omega_{y{\tilde x}} - 1 ) dx \wedge d{\tilde x} \wedge dy \wedge 
d{\tilde y}\,,   \\
& = & \lambda^2 (\{\Omega_x,\Omega_y\} - 1 ) dx \wedge d{\tilde x} \wedge dy \wedge 
d{\tilde y}\,,   \\
& = & 0 
\end{eqnarray*}

\noindent by virtue of (\ref{eq:mPlebanski}). Thus the Lax pair, and hence the 
integrability of this deformed system is encoded into the equations

\begin{eqnarray*}
d{\bf\Omega} & = & 0 \,, \\
{\bf\Omega}\wedge{\bf\Omega} & = & 0 \,.
\end{eqnarray*}

\bigskip

Further properties of this system have been found. In [S92] a perturbative 
solution in powers of $\kappa$ was constructed. On writing

\[
\Omega=\sum_{n=0}^\infty \kappa^n \Omega_n
\]

\noindent one obtains Plebasnki's equation (\ref{eq:Plebanski}) for $\Omega_0$ 
and an infinite number of linear equations for the $\Omega_n\,,n>0$ of the form

\[
\Box_{\Omega_0} \Omega_n = S_n[\Omega_0\,,\ldots\,,\Omega_{n-1}]\,,\quad\quad 
n=1\,,2\,,\ldots\,,\infty\,.
\]

\noindent The operator $\Box_{\Omega}$ is the wave operator on the spacetime 
given by the metric ${\bf g}(\Omega)$ given by equation (\ref{eq:metric}) and 
the
function $S_n$ is some known function of its arguments. 
A similar procedure may be applied to a Moyal-algebraic deformation of 
Plebanski's $2^{\rm nd}$ heavenly equation [PPRT].
In [C] a symmetry 
reduction of this system was studied and in [T94a] the dressing transform
(using a Riemann-Hilbert factorisation in the corresponding Moyal loop group)
was constructed. As mentioned earlier, one may study the conservation laws, 
symmetries and hierarchies associated with Plebanski's equation and these 
results still holds, under the replacement of the Poisson bracket by the Moyal 
bracket, for the deformed system (\ref{eq:mPlebanski}).

\bigskip

A slightly more general framework may be achieved by observing that the vector 
fields $\partial_x\,,\partial_y\,,X_f$ and $X_g$ which make up the vector 
fields $\cal U$ and $\cal V$ all preserve the volume form 
$\omega=dx\wedge dy\wedge d{\tilde{x}}\wedge d{\tilde{y}}$ on $\M\,,$ as in the 
construction of self-dual metrics [MN].

\bigskip

\subsection*{3.2 The KP hierarchy}

\bigskip

The KP hierarchy is defined as follows. Let ${\cal L}$ be the 
pseudo-differential operator

\[
{\cal L} = \partial + \sum_{n=1}^\infty u_n(x,{\bf t}) \partial^{-n}
\]

\noindent where ${\bf t}=\{t_1\,,t_2\,,\ldots\}\,.$ The evolution of the fields
$u_n(x,{\bf t})$ with respect to the times $\bf t$ is given by the Lax 
equations

\begin{equation}
{\partial{\cal L}\over\partial t_n} = [{\cal B}_n,{\cal L}]
\label{eq:KPLax}
\end{equation}

\noindent where

\[
{\cal B}_n = [{\cal L}^n]_{+} \,, \quad\quad n=1\,,2\,,\ldots\,,\infty
\]

\noindent and $[{\cal O}]_{+}$ denotes the projection onto the purely 
differential part of the pseudo-differential operator ${\cal O}\,.$ Similarly,
$[{\cal O}]_{-}$ denotes the projection onto purely negative powers, so
${\cal O} = {\cal O}_{+} + {\cal O}_{-}\,.$

\bigskip

Let the coordinates on $\M$ be $\{x,\lambda\}$ and times $\bf t$ be coordinates 
on ${\cal T}\,.$ The $\star$ and $\circ$ products on $\M$ will be given by
equations (\ref{eq:mstar}) and (\ref{eq:mcirc}) evaluated at $\kappa=1$, and    
the ${}^\prime$ will be 
dropped on the corresponding bracket for notional convenience. Taking the 
symbols of the above operators gives the following functions on 
$\widetilde\M\,:$

\begin{eqnarray*}
L & = & sym({\cal L}) \,, \\
  & = & \lambda + \sum_{n=1}^\infty u_n(x,{\bf t}) \lambda^{-n} \,, \\
B_n & = & sym({\cal B}_n)\,. 
\end{eqnarray*}

\noindent Note that

\[
sym({\cal L}^n) = sym({\cal L}) \star_{\kappa=1}  \ldots \star_{\kappa=1} 
sym(\cal L)\,, 
\]

\noindent (which does not equal $sym({\cal L})^n)$
which Kupershmidt [K90] denotes by $L^{\star n}\,$ (see also [F-FMR]).
The Lax equation (\ref{eq:KPLax}) becomes the vector field 
equation

\begin{equation}
L_n(L) = 0 \,,\quad\quad\quad n=1\,,2\,,\ldots\,,\infty\,.
\label{eq:mKPLax}
\end{equation}

\noindent Here $L_n\in T{\widetilde\M}$ is the vector field

\[
L_n = {\partial\phantom{t_n}\over\partial t_n} - X_{B_n} 
\]

\noindent  and the operations are performed with $\kappa=1\,.$ At this point 
this condition will be dropped, thus obtaining a $\kappa$-dependent 
KP hierarchy. The Lax function $L$ remains unchanged, but the $B_n$ 
acquire $\kappa$ dependence, since they are now 
defined by the formula

\[
B_n = [L^{\star n}]_{+}
\]

\noindent (and so reduce to the previous definition if $\kappa=1$), where 
${_{+}}$ 
denotes the projection onto non-negative powers of $\lambda,.$ This 
has the advantage that one may recover the dispersionless KP hierarchy in the
$\kappa\rightarrow 0$ limit without the need for rescaling variables.

\bigskip

\noindent{\bf Example 4}

\bigskip

The first few equations in the $\kappa$-dependent KP hierarchy
(\ref{eq:mKPLax})~are [K90]

\begin{eqnarray*}
B_1 & = & \lambda \,, \\
B_2 & = & \lambda^2 + 2 u_2 \,, \\
B_3 & = & \lambda^3 + 3 \lambda u_1 + 3 u_2 + 3\kappa u_{1,x} \\
\end{eqnarray*}

\noindent which leads to the evolution equations

\begin{eqnarray*}
u_{1,t_2} & = & 2 u_{2,x} + \kappa u_{1,xx} \,, \\
u_{2,t_2} & = & 2 u_{3,x} + 2 u_1 u_{1,x} + \kappa u_{2,xx} \,, \\
u_{1,t_3} & = & 3 u_{3,x} + 6 u_1 u_{1,x} + 3\kappa u_{2,xx} + \kappa^2 
u_{1,xxx} 
\end{eqnarray*}

\noindent (the $t_1$-flows are trivial). These show the $\kappa$-dependent 
terms. On eliminating $u_2$ and $u_3$ one 
obtains the KP equation

\[
(4u_{1,t_3} - 12 u_1 u_{1,x} - \kappa^2 u_{1,xxx})_x = 3 u_{1,t_2 t_2}\,.
\]

\noindent One may obtain an equivalent KP hierarchy by using the Moyal 
$\star$-product and bracket rather than the Kupershmidt-Manin $\star$-product  
and bracket [K,S95a]. One obtains the functions 

\begin{eqnarray*}
B_1 & = & \lambda \,, \\
B_2 & = & \lambda^2 + 2 u_1 \,, \\
B_3 & = & \lambda^3 + 3 \lambda u_1 + 3 u_2 
\end{eqnarray*}

\noindent ($\kappa$-dependent terms only appear in the $B_n$ for $n>3$) and 
evolution equations 

\begin{eqnarray*}
u_{1,t_2} & = & 2 u_{2,x} \,, \\
u_{2,t_2} & = & 2 u_1 u_{1,x} + 2 u_{3,x} \,, \\
u_{1,t_3} & = & 6 u_1 u_{1,x} + 3 u_{3,x} + {\kappa^2\over 4} u_{1,xxx} \,. 
\end{eqnarray*}

\noindent This system also leads to the KP equation on eliminating $u_2$ and 
$u_3\,,$ and on redefining the fields it is easy to see that these two 
systems are equivalent.
Note that in both cases the limit $\kappa\rightarrow 0$ one obtains the 
dispersionless KP equation directly without further rescaling of the variables.

\bigskip

\endexample

\bigskip

An alternative form of the KP hierarchy is based on the zero-curvature 
conditions (which follow from the Lax equation (\ref{eq:KPLax}))

\[
{\partial{\cal B}_n\over\partial t_m}-{\partial{\cal B}_m\over\partial t_n} +
[{\cal B}_n,{\cal B}_m] = 0
\]

\noindent or equivalently

\begin{equation}
{\partial{B}_n\over\partial t_m}-{\partial{B}_m\over\partial t_n}
+\{{B}_n,{B}_m\} = 0 \,.
\label{eq:KPzerocurv}
\end{equation}

\noindent Note that this is the condition of the vector fields $L_n$ to 
commute, $[L_m,L_n]=0$ for all $m\,,n=1\,,2\,,\ldots\,,\infty\,.$

\bigskip

These zero-curvature relations may be encoded in a 2-form $\bf\Omega$ defined by

\[
{\bf\Omega} = d\lambda\wedge dx + \sum_{n=2}^\infty d B_n \wedge d t_n\,.
\]

\noindent This form satisfies the following equations

\begin{eqnarray*}
d{\bf\Omega} & = & 0 \,,   \\
{\bf\Omega}\wedge{\bf\Omega} & = & 0 \,.
\end{eqnarray*}

\noindent The first is obvious. One has to be careful in evaluating the second 
equation (see Example 2 in section 2), but one obtains

\[
{\bf\Omega}\wedge{\bf\Omega} = \sum_{m,n=2}^\infty
\Big[
{\partial{B}_n\over\partial t_m}-{\partial{B}_m\over\partial t_n}
+\{{B}_n,{B}_m\}
\Big] d\lambda \wedge dx \wedge d t_m \wedge d t_n
\]

\noindent and hence ${\bf\Omega}\wedge{\bf\Omega} = 0$ if and only if the zero-curvature 
relations (\ref{eq:KPzerocurv})~hold. The geometry of the KP hierarchy will be 
discussed further in section 4.

\bigskip

\subsection*{3.3 The Toda hierarchy}

\bigskip

The definition of this hierarchy is very similar to the definition of the KP 
hierarchy. The Lax operator is

\[
{\cal L} = e^{\partial} + \sum_{n=0}^\infty u_n(x,{\bf t}) e^{-n\partial}
\]

\noindent (note the range of summation) and the evolutions of the fields are 
given by the Lax equation

\[
{\partial{\cal L}\over\partial t_n} = [ {\cal B}_n , {\cal L}]
\]

\noindent where

\[
{\cal B}_n = [{\cal L}^n]_{+} \,,\quad\quad n=1\,,2\,,\ldots\,,\infty
\]

\noindent and $[{\cal O}]_{+}$ denotes the projection onto positive powers of
$e^\partial$ of the pseudo-differential operator ${\cal O}\,.$ The operator
$e^\partial$ acts as a shift operator,

\[
e^{n\partial} f(x) = f(x+n)\,.
\]

The geometric description of the hierarchy is obtained in the same way as 
above. Taking symbols of the operators give

\begin{eqnarray*}
L & = & sym({\cal L}) \,, \\
  & = & e^\lambda + \sum_{n=0}^\infty e^{-n\lambda} \,, \\
B_n & = & sym({\cal B}_n)
\end{eqnarray*}

\noindent and the above Lax equation becomes

\begin{equation}
{\partial L\over\partial t_n} = \{ B_n , L\}\,.
\label{eq:TodaLax}
\end{equation}

\noindent Once again the condition $\kappa=1$ will be dropped, so composition 
will be done using the Kupershmidt-Manin $\star$- and $\circ$-products 
(\ref{eq:mstar}) and (\ref{eq:mcirc}), so now the $B_n$ are defined by the 
equation $B_n=[L^{\star n}]_{+}$ where ${_{+}}$ denotes the projection onto 
non-negative powers of $e^\lambda\,,$ as in the $\kappa$-dependent KP hierarchy. 
In the limit $\kappa\rightarrow 0$  
one obtains the dispersionless Toda hierarchy. One difference between the 
hierarchy and the KP hierarchy is that the evolution equations for the fields 
contains an infinite number of $\kappa$-dependent terms. However these may be 
recombined in terms of shift operators, as the following example will show.

\bigskip

\noindent{\bf Example 5}

\bigskip

One possible truncation of this hierarchy is to set $u_n=0$ for $n\geq 2\,,$ so 
that

\begin{eqnarray*}
L & = & e^\lambda + u_0 + u_1 e^{-\lambda} \,, \\
B_1 & = & e^\lambda + u_0,.
\end{eqnarray*}

\noindent The evolution equations for the fields $u_0$ and $u_1$ are given by

\[
{\partial L\over \partial t} = \{ B_1 , L \}
\]

\noindent where the bracket is the Kupershmidt-Manin bracket (\ref{eq:KMB}) 
and, for greater generality, the $\kappa=1$ condition has been dropped. 

This gives the equations

\begin{eqnarray*}
u_{0,t}(x) & = & {1\over\kappa} \Bigg[ \sum_{s=0}^\infty {\kappa^s\over s!}
\partial_x^s \, - 1 \, \Bigg] \, u_1(x) \,, \\
& & \\
& = & {u_1(x+\kappa) - u_1(x)\over \kappa} \,, \\
& & \\
u_{1,t}(x) & = & {u_1(x)\over\kappa} \Bigg[ 1 - \sum_{s=0}^\infty 
{(-1)^s \kappa^s \over s!} \partial_x^s \Bigg] \, u_0(x) \,, \\
& & \\
& = & {u_1(x) [ u_0(x) - u_0(x-\kappa) ]\over \kappa }
\end{eqnarray*}

\noindent and on eliminating $u_0$ on obtains the Toda lattice equation

\[
(\log u_1(x)\, )_{tt} = 
{u_1(x+\kappa) - 2 u_1(x) + u_1(x-\kappa)\over\kappa^2}\,.
\]

Instead of using the Kupershmidt-Manin bracket one could use the Moyal 
bracket. This gives the slightly different equations

\begin{eqnarray*}
u_{0,t}(x) & = & {2\over\kappa} \Big[ \sum_{s=0}^\infty 
{\kappa^{2s+1}\over 2^{2s+1} (2s+1)!} \partial_x^{2s+1} \Big] u_1(x) \,, \\
& & \\
& = & {1\over\kappa}\Big[u_1(x+\kappa /2)-u_1(x-\kappa /2)\Big] \,, \\
& & \\
u_{1,t}(x) & = & {2u_1(x)\over\kappa} \Big[ \sum_{s=0}^\infty
{\kappa^{2s+1}\over 2^{2s+1} (2s+1)!} \partial_x^{2s+1} \Big] u_0(x) \,, \\
& & \\
& = & {u_1\over\kappa}\Big[u_0(x+\kappa /2)-u_0(x-\kappa /2)\Big] \,.
\end{eqnarray*}

\noindent On eliminating $u_0$ one again recovers the Toda lattice equation.
Note that with the Kupershmidt-Manin bracket one obtains a forward/backward 
difference operator while the Moyal bracket gives a central difference 
operator.
In both cases $\kappa$ acts as the lattice spacing and 
as $\kappa\rightarrow 0$ one recovers the dispersionless Toda 
equations (since both these brackets are deformations of the Poisson bracket)

\begin{eqnarray*}
u_{0,t} & = &  u_{1,x} \,, \\
u_{1,t} & = &  u_1 u_{0,x}\,.
\end{eqnarray*}

\noindent Some of the properties of this system and its hierarchy may be found 
in [K85,FS].

\bigskip

\endexample

\bigskip

As with the KP hierarchy, the Lax equation (\ref{eq:TodaLax}) is 
equivalent to a set of zero-curvature relations for the $B_n$ and these may be 
encoded into a 2-form $\bf\Omega$ which satisfies the equations

\begin{eqnarray*}
d{\bf\Omega} & = & 0 \,, \\
{\bf\Omega}\wedge{\bf\Omega} & = & 0
\end{eqnarray*}

\noindent in exactly the same way as was done 
for the KP hierarchy.

\bigskip

\section*{4. The Geometry of the KP hierarchy}

\bigskip

The main result of this section is to show how a solution of the KP hierarchy 
may be associated to a solution of a Riemann-Hilbert problem in the Lie group 
$SDiff_\kappa(\M^2)$ (the Lie group corresponding to the Lie algebra 
$sdiff_\kappa(\M^2)\,$. Explicitly, given a map

\[
\left( \begin{array}{c} x \\ k \end{array} \right) \mapsto
\left( \begin{array}{c} f(x,k) \\ g(x,k) \end{array} \right) 
\]

\noindent with $\{f,g\}=1$ then this map factors, so there exists a map

\[
\left( \begin{array}{c} P \\ Q \end{array} \right) \mapsto
\left( \begin{array}{c} f(P,Q) \\ g(P,Q) \end{array} \right) 
\]

\noindent where the right hand side is analytic in $k$ (the notation 
$S_{-}$
will be used to denote the part of the Laurent series $S$ consisting of 
negative powers of $k$ only). The results derived in section 2 enable existing 
results on the KP hierarchy to be lifted whilst furnishing them with a 
geometrical interpretation (the definitions of the manifold $\widetilde{\M}$ 
and $\star$- and $\circ$-products 
will be the same as in section 3.2). In this section $\kappa=1$ and the $\kappa$-symbol 
on the exponential $\exp_\kappa$ will be dropped. The main results of this 
section are due to Takasaki and Takebe [TT]. A more careful analysis is needed for
$\kappa\neq 1\,.$

\bigskip

More fundamental than the Lax operator $\cal L$ is the operator $\cal W$ 
defined by

\[
{\cal W} = 1 + \sum_{n=1}^\infty w_n \partial^{-n}
\]

\noindent with which the Lax operator is defined as

\[
{\cal L} = {\cal W} \partial {\cal W}^{-1}\,.
\]

\noindent The evolution of the fields $w_n$ are given by the equation

\[
{\partial {\cal W} \over \partial t_n} = {\cal B}_n {\cal W} - {\cal W} 
\partial^n
\]

\noindent from which follows the Lax equation (\ref{eq:KPLax}). On taking the 
symbols of the operators one obtains

\begin{eqnarray*}
W & = & sym({\cal W}) \,, \\
  & = & 1 + \sum_{n=1}^\infty w_n k^{-n} \,, \\
  &   &    \\
L & = & sym({\cal L}) \,, \\
  & = & Ad(W) \, k \,.
\end{eqnarray*}

\noindent The Orlov operator $\cal M$ is defined by [GO]

\[
{\cal M} = {\cal W} ( \sum_{n=1}^\infty n t_n \partial^{-n} + x )  
{\cal W}^{-1} 
\]

\noindent or, equivalently, by

\begin{eqnarray*}
M & = & sym({\cal M}) \,, \\
  & = & Ad(W \exp[{\bf t}(k)]\,) x
\end{eqnarray*}

\noindent where ${\bf t}(k) = \sum_{n=1}^\infty t_n k^n\,.$

\bigskip

\noindent {\bf Lemma}

\bigskip

The pair $(L,M)$ satisfy the equations

\begin{eqnarray*}
{\partial L\over \partial t_n} & = & \{ B_n , L \} \,, \\
{\partial M\over \partial t_n} & = & \{ B_n , M \} \,, \\
\{ L , M \} & = & 1 \,.
\end{eqnarray*}

\noindent Conversely, given such a pair then there exist a unique dressing 
function $W$ so that $L=Ad(W) k$ and $M=Ad(W \exp[{\bf t}(k)]\,) x\,.$

\endproof

\bigskip

Such a $(L,M)$ pair will be said to satisfy the KP hierarchy.
The first two of these equations may be 
re-written as vector field equations: 

\begin{eqnarray*}
( \partial_{t_n} - X_{B_n} ) \,L & = & 0 \,, \\
( \partial_{t_n} - X_{B_n} ) M & = & 0 \,,
\end{eqnarray*}

\noindent or alternatively, using the inner product $\langle , \rangle$ between
vector fields and forms, as

\begin{eqnarray*}
\langle \partial_{t_n} - X_{B_n} , dL \rangle & = & 0 \,, \\
\langle \partial_{t_n} - X_{B_n} , dM \rangle & = & 0 \,
\end{eqnarray*}

\noindent for $n=1\,,2\,,\ldots\,,\infty\,.$
These functions 
$(L,M)$ will play the analogous roles to the coordinates on the twistor 
surfaces in the nonlinear gravition construction.

\bigskip

The next theorems show how such a pair are related to a Riemann-Hilbert 
factorization problem. The first show how a solution to the Riemann-Hilbert 
problem defines a solution to the KP hierarchy and the second shows the 
converse. 

\bigskip

\noindent{\bf Theorem}

\bigskip

Suppose on has functions

\begin{eqnarray*}
L & = & Ad(W) k \,,    \\
M & = & Ad(W \exp[{\bf t}(k)]\,) x
\end{eqnarray*}

\noindent (with $\{L,M\}=1\,$). Then for any pair of functions 
$f(x,k)\,,g(x,k)$ (with Laurent series) if

\begin{eqnarray*}
\{ f , g \} & = & 1 \,, \\
f(M,L)_{-} & = & 0 \,, \\
g(M,L)_{-} & = & 0 \,, 
\end{eqnarray*}

\noindent then the pair $(L,M)$ satisfies the KP hierarchy.

\endproof

\bigskip

\noindent{\bf Theorem}

\bigskip

If the pair $(L,M)$ satisfy the KP hierarchy then there exist functions $f\,,g$ 
such that

\begin{eqnarray*}
\{ f , g \} & = & 1 \,, \\
f(M,L)_{-} & = & 0 \,, \\
g(M,L)_{-} & = & 0 \,. 
\end{eqnarray*}

\endproof

\bigskip

The proofs are basically identical to the proofs in [TT], the only difference 
being that here they are reformulated in terms of the deformed geometric 
structures rather than in terms of pseudo-differential operators. One may also 
prove an uniqueness results for the solution $(L,M)$, at least in the 
neighbourhood of the trivial solution $(k,x)\,.$

\bigskip

One outstanding problem is how to relate the $2$-form $\bf\Omega$ to the pair 
$(L,M)\,.$ In the dispersionless limit one has (for all the systems discussed 
in section $3$) a relation

\[
{\bf\Omega} = d L \wedge d M\,.
\]

\noindent However, the proof of this result relies on the associative property 
of normal multiplication which nolonger holds for the deformed $\circ$-product.
It may be that this result still holds, for example, the obstruction might 
vanish due to the relation $\{L,M\}=1\,.$ This problem, of how to understand 
the direct relation between the pair $( L,M)$ and $\bf\Omega$ is currently under 
investigation.

\bigskip

\section*{5. Comments}

\bigskip

In summary, the three classes of integrable hierarchy discussed in section $3$ 
may all be formulated in terms of vector fields ${\cal V}_i$ which 
preserve a volume form

\[
\omega=dx \wedge dy \wedge \bigwedge_{n=1}^\infty dt_n
\]

\noindent (or $\omega=dx \wedge dy \wedge d{\tilde{x}} \wedge d{\tilde{y}}$ in 
the deformed anti-self-dual vacuum equations) on $\M\,,$ together with function 
$(L,M)$ which satisfy the equations

\begin{eqnarray*}
{\cal V}_i (L) & = & 0 \,, \\
{\cal V}_i (M) & = & 0 \,, 
\end{eqnarray*}

\noindent or, in terms of the inner product between $T\M$ and $T^{*}\M\,,$

\begin{eqnarray*}
\langle {\cal V}_i , dL \rangle & = & 0 \,, \\
\langle {\cal V}_i , dM \rangle & = & 0 \,, \\
\end{eqnarray*}

\noindent with the functions $L$ and $M$ being independent: $\{L,M\}=1\,.$
A dual description also exists for all these systems in terms of a two form 
$\Omega$ on $\M$ satisfying the equations

\begin{eqnarray*}
d \Omega & = & 0 \,, \\
\Omega \wedge \Omega & = & 0 \,.
\end{eqnarray*}

\noindent The precise relationship between these two dual descriptions requires 
further investigation. In all cases the solutions are encoded in a 
Riemann-Hilbert problem in the corresponding loop group.

\bigskip

This work raises a number of further question. For example, it should be 
straightforward to examine the sysmmetries of these integrable systems using 
thses methods, with the Hamiltonian vector fields playing the role of the 
symmetry generators (see, for example [T94b]). 
This would provide a geometrical description of various 
$W_\infty$ and $W_{\rm KP}$ algebras. One use of such symmetries is in the 
construction of a constrained KP-hierarchy. One example of this contains the 
KdV hierarchy. However, the KdV hierarchy has been shown to be a reduction of 
the self-dual Yang-Mills equations (and its generalisations). Thus there are 
two way of looking at the KdV equation: one based on the Yang-Mills 
self-duality equation and one based on the deformed differential geometry 
constructed in section 2. Precisely how these two seemingly different 
constructions are related deserves further study. In connection with this is 
how to understand the non-local nature of the Riemann-Hilbert problem for the 
KP equation compared with the local one for the KdV equation [AC,M].

\bigskip

All of the examples of integrable systems studied in this paper have used 
Hamiltonian vector fields in their construction. Are there any systems which 
use more general, non-Hamiltonian, vector fields? The property of commuting
flows for these hierarchies can be traced back to the Jacobi identity for 
Hamiltonian vector fields, so any hierarchy based on non-Hamiltonian vector 
fields might lose this property.

\bigskip

One possible use of this deformed calculus would be to develop a theory of 
deformed (or quantum) twistor spaces (which would encode the Riemann-Hilbert 
problems in the Moyal loop group) more axiomatically. One obvious place to 
start is to deform the sympletic structure on the fibres of the nonlinear 
graviton's twistor space. An observation that might be of use is that 
$\star$-product do exist on the complex manifold ${\bf CP}^3$ and other complex 
coset spaces. This suggest that one should develop a deformation theory (in the 
sense of Kodaira) of such spaces. Such idea, however, are outside the scope of 
this paper.

\bigskip

\section*{Acknowledgements}

I would like to thank the University of Newcastle, where this paper was
written, for the Wilfred Hall 
Fellowship and Kanehisa Takasaki and David Fairlie for their comments on this work.

\bigskip

\section*{Notes added in proof}

Since this paper was written a number of other papers have appeared. In
[PP] and [G-CPP] the Moyal deformation of self-dual gravity has been studied
using a chiral model approach and in [S96] it was shown that the Toda lattice is
a reduction of this Moyal deformed self-dual gravity, a result analogous to
the reduction from the standard, undeformed, self-dual gravity equations to the
Boyer-Finley equation. Other notable papers are [Ke], [KeS],[DM-H] and [R],
which develop various connections between discrete systems, geometry and
associative $\star$-products.

\section*{References}

\bigskip
\begin{tabbing}
References \=
\kill
\noindent{[A]} \> Averson, W., Quantization and uniqueness of invariant 
structures, \\
\> Commun. Math. Phys. {\bf 89} (1983) 77-102.\\

\\
\noindent{[AC]} \> Ablowitz, M.J. and Clarkson, P.A., {\sl Solitons, non-linear 
evolution equations} \\
\> {\sl and inverse scattering}, LMS Lecture Note Series 149, CUP, Cambridge. \\

\\
\noindent{[BFFLS]} \> Bayen, F., Flato, M., Fronsdal, C., Lichnewowicz, A. and
Sternheimer, D.,\\
\> {\sl Deformations as quantisations, I,II}, Ann. of Phys. NY {\bf 111} (1978)
61-110, 111,151. \\

\\
\noindent{[C]} \> Castro, C., {\sl Nonlinear $W_\infty$ algebras from nonlinear 
integrable deformations} \\ \> {\sl of self dual gravity}, Preprint IAEC-4-94, \\ \> {\sl A Universal $W_\infty$ algebra and quantization of 
integrable deformations} \\ \> {\sl of self-dual gravity}, Preprint IAEC-2-94.\\

\\
\noindent{[DM-H]} \> Dimakis, A and M\"uller-Hoissen, F. {\sl Integrable discretizations of
chiral models}\\
\>{\sl via deformations of the differential calculus}, preprint hep-th/9512007. \\

\\
\noindent{[Fe]} \> Fedosov, B., {\sl A simple geometrical construction of deformed quantization},\\
\> J. Diff. Geom. {\bf 140} (1994) 213-238.\\

\\
\noindent{[Fl]} \> Fletcher, P., {\sl The uniqueness of the Moyal algebra}, \\
\> Phys. Lett. {\bf B248} (1990) 323 - 328. \\

\\
\noindent{[F-FMR]} \> Figueroa-O'Farrill, J.M., Mas, J. and Ramos, E., {\sl A 
One-Parameter Family}\\ \> {\sl of Hamiltonian Structures for the KP Hierarchy and a 
Continuous }\\ \> {\sl Deformation of the Nonlinear ${\rm W}_{\rm KP}$ Algebra,}\\ \> Commun. Math. Phys. {\bf 158} (1993) 17-43. \\

\\
\noindent{[FS]} \> Fairlie, D.B. and Strachan, I.A.B., {\sl The Hamiltonian structure of 
the} \\ 
\> {\sl dispersionless Toda hierarchy}, Physica D {\bf 90} (1996) 1 - 8.\\

\\
\noindent{[G]} \> Gindikin, S.G., {\sl Generalized Conformal Structures}, in \\
\> {\sl Twistors in Mathematics and Physics}, ed. Bailey, T.N. and Baston, R.J. \\
\> LMS Lecture Note Series 156, CUP, Cambridge. \\

\\
\noindent{[G-CPP]} \> Garcia-Compean, H., Plebanski, J.F. and Przanowski, M, \\
\> {\sl From principal chiral model to self-dual gravity}, \\
\> CINVESTAV-FIS-17/95, hep-th/9509092 \\
\> {\sl Further remarks on the chiral models approach to self-dual gravity}, \\
\> CINVESTAV-FIS-22/95, hep-th/9512013.\\

\\
\noindent{[GO]} \> Grinevich, P.G. and Orlov, A,Yu., {\sl Virasoro action on Riemann surfaces,} \\ \> {\sl Grassmannians, $\det{\overline{\partial}}_J$ and Segal-Wilson $\tau$-function}, \\
\> in {\sl Problems of Modern Quantum Field Theory},\\ 
\> ed. A.A.Belavin et.al. (Berlin: Springer, 1989). \\

\\
\noindent{[K85]} \> Kupershmidt, B.A., {\sl Discrete Lax Equations and 
Differential-difference} \\
\> {\sl Calculus}, Asterisque, {\bf 123} (1985) 1-212. \\

\\
\noindent{[K90]} \> Kupershmidt, B.A., {\sl Quantizations and Integrable Systems}, \\
\> Lett. Math. Phys. {\bf 20} (1990) 19-31. \\

\\
\noindent{[Ke]} \> Kemmoku, R., {\sl Phase space discretization and its application }\\
\> {\sl to integrable systems}, preprint. \\

\\
\noindent{[KeS]} \> Kemmoku, R. and Saito, S., {\sl Phase space discretization and}\\
\> {\sl Moyal quantization}, TMUP-HEL-9507, hep-th/9510077.\\

\\
\noindent{[M]} \> Mason, L.J. {\sl Generalized Twistor Correspondences, D-Bar 
Problems and} \\
\> {\sl the KP Equations}, in {\sl Twistor Theory}, ed. Huggett, S., Lecture    
notes in \\
\> pure and applied mathematics vol 169 (New York: Dekker).\\

\\
\noindent{[Ma]} \> Manin, Yu.,I, J. Sov. Math. {\bf 11} (1979) 1-122.\\

\\
\noindent{[MN]} \> Mason, L.J. and Newman, E.T., {\sl A connection between the 
Einstein and} \\
\> {\sl Yang-Mills equations}, Commun. Math. Phys. {\bf 121} (1989) 659 - 
668.\\

\\
\noindent{[Mo]} \> Moyal, J., {\sl Quantum Mechanics as a Statistical Theory}, 
\\ 
\> Proc. Camb. Phil. Soc. {\bf 45} (1949) 99 - 124. \\

\\
\noindent{[MS]} \> Mason, L,J. and Sparling, G.A.J., {\sl Nonlinear Schr\"odinger and 
Korteweg} \\
\> {\sl deVries are reductions of self-dual Yang-Mills},\\
\> Phys. Lett. {\bf A 137} (1989) 29-33. \\
\> {\sl Twistor correspondences for the soliton hierarchies}, \\
\> J. Geom. Phys. {\bf 8} (1992) 243-271. \\

\\
\noindent{[NPT]} \> Newman, E.T., Porter, J.R. and Tod, K.P. {\sl Twistor 
surfaces and right-flat} \\ 
\> {\sl spaces}, Gen. Rel. Grav. {\bf 9} (1978) 1129. \\

\\
\noindent{[Pa]} \> Park, Q-H, {\sl Self-dual gravity as a large-N limit of the 2D 
non-linear sigma} \\
\> {\sl model}, Phys. Lett. {\bf B238} (1990) 287-290.\\
\> {\sl 2-D sigma model approach to 4-D instantons},\\
\> Int. J. Mod. Phys. {\bf A7} (1992) 1415-1448.\\

\\
\noindent{[Pen]} \> Penrose, R. {\sl Nonlinear gravitions and curved twistor theory},\\
\> Gen. Rel. Grav. {\bf 7} (1976) 31-52 \\
\> {\sl The nonlinear Graviton}, Gen. Rel. Grav. {\bf 7} (1976) 171-176.\\

\\
\noindent{[Pl]} \> Plebanski, J.F., {\sl Some solutions of complex Einstein 
equations}, \\
\> J. Math. Phys. {\bf 16} (1975) 2395-2402. \\

\\
\noindent{[PP]} \> Plebanski, J.F.and Przanowski, M., {\sl The Lagrangian of a
self-dual}\\
\> {\sl gravitional field as a limit of the SDYM Lagrangian,} \\
\> Phys. Lett. {\bf A212} (1996) 22 - 28.\\

\\
\noindent{[PPRT]} \> Plebanski, J.F., Przanowski, M., Rajca, B. and Tosiek, K,\\
\>{\sl The Moyal deformation of the second heavenly equation}, \\ 
\>{\sl Acta Physica Polonica} {\bf B26} (1995) 889-902. \\

\\
\noindent{[R]} \> Reuter, M., {\sl Non-commutative geometry on quantum phase space,}\\
\> preprint hep-th/9510011. \\

\\
\noindent{[S92]} \> Strachan, I.A.B., {\sl The Moyal algebra and integrable 
deformations of the} \\
\> {\sl self-dual Einstein equations}, Phys. Lett. {\bf B283} (1992) 63-66. \\

\\
\noindent{[S93]} \> Strachan, I.A.B., {\sl Hierarchy of Conservation Laws for 
self-dual gravity},  \\
\> Class. Quantum. Grav. {\bf 10} (1993) 1417-1423. \\

\\
\noindent{[S95a]} \> Strachan, I.A.B., {\sl The Moyal bracket and the dispersionless 
limit of the}\\
\> {\sl KP hierarchy}, J. Phys. A. {\bf 20} (1995) 1967-1975.\\

\\
\noindent{[S95b]} \> Strachan, I.A.B., {\sl The symmetry structure of the 
anti-self-dual Einstein }\\
\> {\sl hierarchy}, J. Math. Phys. {\bf 36} (1995) 3566-3573. \\

\\
\noindent{[S96]} \> Strachan, I.A.B., {\sl The dispersive self-dual Einstein equations}\\
\> {\sl and the Toda Lattice}, preprint. \\

\\
\noindent{[T94a]} \> Takasaki, K., {\sl Dressing operator approach to Moyal algebraic 
deformations} \\
\> {\sl of self-dual gravity}, J. Geom. Phys. {\bf 14} (1994) 111-120, \\
\> {\sl Non-Abelian KP hierarchy with Moyal algebraic coefficients},\\
\> J. Geom. Phys. {\bf 14} (1994) 332-364. \\
\\
\noindent{[T94b]} \> Takasaki, K., {\sl Symmetries and tau function of higher 
dimensional} \\ 
\> {\sl dispersionless integrable hierarchies}, J. Math. Phys. {\bf 36} (1995)
3574-3607. \\

\\
\noindent{[TT]} \> Takasaki, K. and Takebe, T., {\sl SDiff(2) KP hierarchy}, \\
\> Int. J. Mod. Phys. {\bf A7} Suppl. 1 (1992) 889-992\\
\> {\sl Integrable hierarchies and dispersionless limit}, \\
\> hep-th 9405096, preprint UTMS 94-35.\\

\\
\noindent{[V]} \> Vasiliev, M.A., {\sl Algebraic aspects of the higher-spin problem}, \\
\> Phys. Lett. {\bf B257} (1991) 111-118,\\
\> {\sl More on equations of motion for interacting massless fields of all
spins }\\
\> {\sl in $3+1$ dimensions}. Phys. Lett. {\bf B285} (1992) 225-234.\\

\\
\noindent{[W77]} \> Ward, R.S., {\sl On self-dual gauge fields}, \\
\> Phys. Lett. {\bf A61} (1977) 81-82. \\

\\
\noindent{[W]} \> Ward, R.S., {\sl Integrable and solvable systems and relations 
among them}, \\
\> Phil. Trans. Roy. Soc. {\bf A315} (1985) 451-457, \\
\> {\sl Multidimensional integrable systems}, in {\sl Field Theory, Quantum Gravity}\\
\>{\sl and Strings}, ed. de Vega, H.J., Sanchez, N., Lecture Notes in Physics 246\\
\> (Berlin, 
Heidelberg, New York: Springer, 1986), \\
\> {\sl Integrable systems in twistor theory}, in {\sl Twistors in Mathematics and 
Physics} \\
\> ed. Bailey, T.N. and Baston, R.J., LMS Lecture Note 156, CUP, Cambridge. \\
\> {\sl Infinite-dimensional gauge groups and special nonlinear gravitons},\\
\> J. Geom. Phys. {\bf 8} (1992) 317-325.\\

\\
\end{tabbing}

\end{document}